\documentclass[12pt]{article} 
\usepackage{graphicx}
\usepackage{bm}

\begin{document}

\title{\bf Polarized Proton Pionic Capture in Deuterium 
as a Probe of $3N$ Dynamics}  

\author{Luciano~Canton$^{1,2}$ and Leonid~G.~Levchuk$^{2,3}$}
\date{}
\maketitle
\begin{center}
{\em 
$^1$ Istituto Nazionale di Fisica Nucleare, 35131 Padova, via Marzolo, n. 8 Italy,\\
$^2$ Dipartimento di Fisica dell'Universit\`a, 35131 Padova, via Marzolo, n. 8 Italy,\\ 
$^3$ Kharkov Institute of Physics and Technology, 61108 Kharkov, Ukraine\\ 
}
\end{center}

\vspace{5mm}

\begin{abstract}
\sloppy
The proton analyzing power $A_y$ in pion production reaction 
$\overrightarrow{p} d \rightarrow \pi^0{\ ^{3}}\rm{He}$ 
has been calculated including one- and two-body meson production mechanisms 
with a proper 
treatment of the three-nucleon dynamics and  
an accurate solution of the 3$N$ bound-state problem for  
phenomenological two-nucleon potentials. In the region around
the $\Delta$ resonance, the structure of the analyzing power can 
be understood once interference effects among amplitudes 
describing intermediate $\Delta N$ formation in different orbital states
are considered along with the additional interference with the $S$-wave 
pion production amplitudes. Then, the inclusion of
three-nucleon dynamics in the initial state produces the
structure of the analyzing power that has been observed experimentally.
\end{abstract}


\maketitle

\newcommand{\ve}[1]{{\bf{#1}}}
\newcommand{\vegr}[1]{{\bm{#1}}}
\fussy
It has been known for almost twenty years that the 
experimental structure of the analyzing power $A_y$ for the 
$\overrightarrow{p} d \rightarrow \pi^0 {\ ^{3}}\rm{He}$ reaction 
(in the region around the $\Delta$ resonance) exhibits a rapid
variation at the c.m. polar angle of $\Theta_{\rm c.m.} =90^{\circ}$, 
with the appearence of an 
additional (second) peak and changing the sign as one moves from 
lower energies to the $\Delta$-resonance region. 
To our knowledge, this behavior of the proton analyzing power has 
never been interpreted theoretically before, while the understanding 
of the underlying mechanisms is of great importance to describe 
and understand the basic $NN\rightarrow NN \pi$ inelasticities in the 
most simple ``complex'' nuclear environment, 
the three-nucleon system. Single pion production represents the first 
process that needs to be understood in order to consider more 
complicated hadronic processes that occur at moderately higher
energy, such as double-pion production, or heavier meson ({e.g.},~$\eta$) 
production. In addition, single pion production 
could also reveal mechanisms involving multinucleon 
exchanges, thus providing important information that bridges 
intermediate-energy phenomena to low-energy three-nucleon and multinucleon
forces, to be used in the study of low-energy nuclear systems.

\sloppy
Theoretically, pion production in proton-deuteron collisions was studied 
long time ago with the deuteron model~\cite{Rud}, which describes the 
$pd \rightarrow \pi^+ t$ reaction in terms of the differential 
cross section for the more elementary $pp \rightarrow \pi^+ d$ process. 
The model has been subsequently refined by Locher and Weber~\cite{LochWeb}, 
who introduced additional mechanisms based on the elastic 
$\pi d \rightarrow \pi d$ cross section, and by Fearing~\cite{Fearing} who gave 
phenomenological estimates of the effects introduced
by optical-model distortions.
The deuteron model has been later extended by Germond and 
Wilkin~\cite{GermWilk} who developed a simple spin-dependent analysis in 
plane waves by which they predicted the deuteron tensor analyzing
power $T_{20}$ to approach its geometrical limit $-\sqrt{2}$ at 
threshold, based on low-energy theorems. Recently, Falk~\cite{Falk} developed 
further the deuteron model in its spin dependent form and performed a 
phenomenological analysis of polarization observables 
near threshold. Calculations including explicitly $\Delta$ dynamics and 
meson-exchange processes were
pioneered in the late '70~\cite{GreeSai}, but the complexity involved in 
the computation of the corresponding amplitudes forced the authors
to restrict themselves with only exploratory calculations. 
Laget and Lecolley~\cite{LagetLec} 
performed calculations that tried to go beyond one- and two-body pion 
production mechanisms and included effective three-body mechanisms 
{via} meson double rescatterings, showing beneficial effects at 
higher energies, well beyond the $\Delta$ resonance. Results around 
the isobar resonance, however, were not quite satisfactory, and 
especially the proton analyzing power $A_y$, {i.e.}, 
the asymmetry of the reaction yield for the opposite proton spin 
directions, 
\begin{equation}
A_y={{N_\uparrow-N_\downarrow}\over {N_\uparrow+N_\downarrow}}\, , 
\end{equation}
was in poor accord with 
the data~\cite{Cameron}. 

\sloppy
The approach involving the $3N$ dynamics has been pushed forward 
recently in a series of 
papers~\cite{CantonSchad}, where explicit $\Delta$ excitation mechanisms
were treated in the three-nucleon system using the few-body 
techniques~\cite{Canton_ea_1}. Intermediate $\Delta$ propagation and 
rescattering were described phenomenologically using an isobar
complex-mass scaling that has been tested and parametrized on the 
$pp \rightarrow \pi^+ d$ reaction~\cite{Canton_ea_2}.
Such test calculations have shown that, for the 
description of the proton pionic capture by
protons, the interplay between ($\Delta$-mediated) $P$-wave mechanisms 
and $S$-wave two-body rescattering mechanisms (in the isoscalar 
and isovector channels) is a fundamental feature that has to be taken 
into account for the overall reproduction of both spin-averaged 
and spin-dependent observables. Anti-symmetrization prescriptions due 
to the identity of the nucleons were fully taken into account.

\sloppy
Nowadays, the calculations can be performed with more 
accurate knowledge of the nuclear wave functions (WF), and with including a 
much larger number of intermediate three-nucleon states. Furthermore, 
the nucleon-deuteron initial-state interaction (ISI) can be calculated 
through, {e.g}, the Alt-Grassberger-Sandhas (AGS) scheme~\cite{AGS}, 
with realistic nucleon-nucleon transition matrices. The complete
calculation of the reaction represents a very complicated theoretical
task, and involves, due to the variety of the few-body intermediate states
that have to be expanded in partial waves, computation of an exorbitant 
number of multidimensional integrals.

\begin{figure}[hbtp]
  \begin{center}
    \resizebox{13cm}{!}{\includegraphics{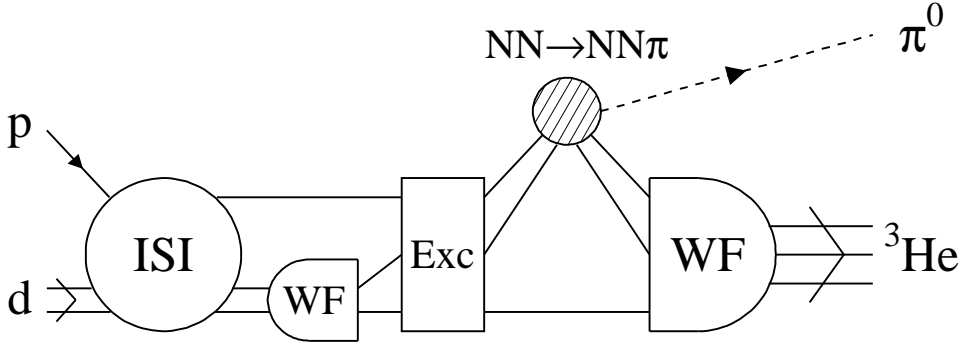}}
\caption{Diagrammatic representation of the pion production amplitude.
}
    \label{diag1}
  \end{center}
\end{figure}

\fussy
The amplitude of the ${p} d \rightarrow \pi^0{\ ^{3}}\rm{He}$ reaction 
(or the inverse pion absorption process) is determined by the matrix element
\begin{equation}
A_{if} = ^{(-)}{_S \langle}\ve{q},\psi_{d} | A |\psi_{t}, \ve{P}^\pi_{0} \rangle\, ,
\label{AMPL}
\end{equation}
where $\psi_t$ represents the three-nucleon bound-state and 
${^{(-)}}{_S\langle}\ve{q},\psi_d |$ refers to a properly antisymmetrized 
interacting nucleon-deuteron state in the initial channel.
Diagrammatically, this amplitude is illustrated by 
Fig.~\ref{diag1}. 
Within the model discussed here, it includes 
the proton-deuteron ISI, the bound-state two- (deuteron) and 
three-nucleon ($^3$He) WF, the matrix element of a $3N$ permutation operator 
that gives rise to the exchange contribution to the amplitude, 
and an operator of pion production on a pair of nucleons.  
Note that in formulation of the AGS formalism employed in our 
calculations, the 3$N$ break-up process is effectively included 
in the ISI block in Fig.~\ref{diag1}.

\begin{figure}[hbtp]
  \begin{center}
    \resizebox{12cm}{!}{\includegraphics{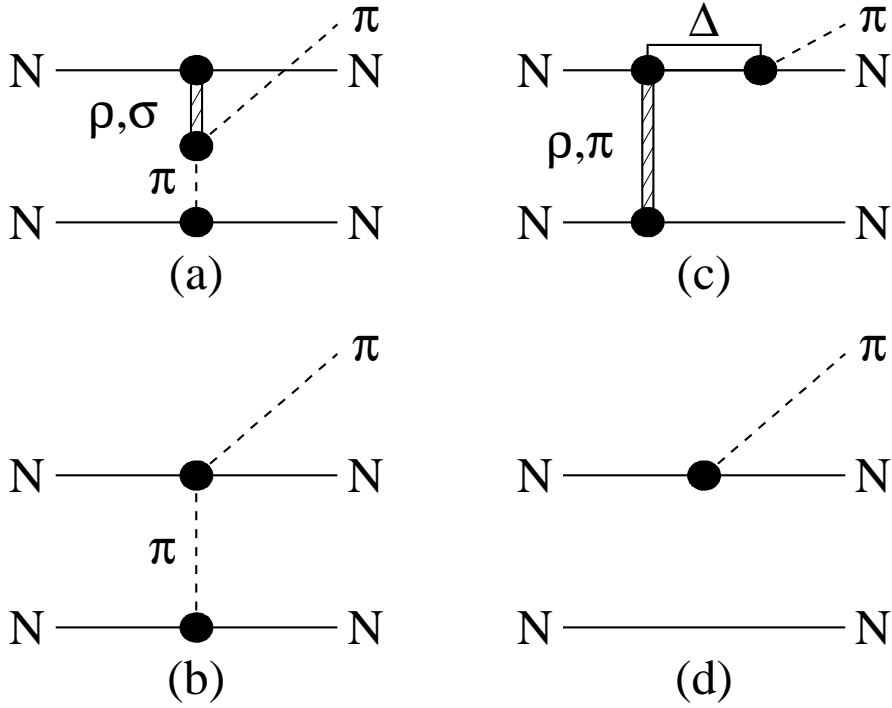}}
\caption{Meson production mechanisms considered.
}
    \label{diag2}
  \end{center}
\end{figure}

 The meson production mechanisms are given by the 
two-body (Fig.~\ref{diag2}a,b,c) 
and one-body  (plain $\pi NN$ vertex in Fig.~\ref{diag2}d) contributions.
The $\Delta$ excitation mechanism (Fig.~\ref{diag2}c) has  the structure 
\begin{equation}
{\cal A}_\Delta = V_{NN-N\Delta} G_\Delta F_{\Delta N\pi} \, ,
\end{equation}
where $V_{NN-N\Delta}$ is the $NN\rightarrow N \Delta$
transition potential generated by $\pi$- and $\rho$-exchanges,
$G_\Delta$ effectively describes the intermediate $\Delta NN$ 
propagation, and $F_{\Delta N\pi}$ is the $\Delta N \pi$ transition 
vertex function. All the corresponding details including the extended formulas 
representing the mechanisms in the three-nucleon partial waves have been 
published elsewhere~\cite{CantonSchad,Canton_ea_1}.
The asymptotic pion wave is not decomposed in partial waves but
treated directly in three dimensions. 
An important quantity for the present calculation is the maximum 
orbital momentum $l_\Delta$ in the $\Delta N$ subsystem. 
We found that $l_\Delta = 0$ is not sufficient to describe 
polarization observables, and one has to go up to $l_\Delta = 2$, 
at least.

In the $\Delta$ resonance region, the spin structure of the reaction amplitude 
is dependent strongly also on the interference effects 
with the other mechanisms, particularly the $S$-wave
two-body pion production contributions shown in Fig.~\ref{diag2}a,b. 
Each of these mechanisms can be schematically described 
as
\begin{equation}
{\cal A}_\pi = F_{N N\pi} G_\pi \lambda_{N\pi-N\pi} + 
{\rm time\ reordering} \ ,
\end{equation}
where $F_{N N\pi}$ is the $\pi N N$ vertex, $G_\pi$ is the pion propagator, and 
$\lambda_{N\pi- N\pi}$ is the corresponding contribution to the half-off-shell 
$\pi N$ scattering amplitude. 
By ``time reordering'' we denote a term coming from 
``backward in time'' propagation of the intermediate-state pion , 
in which case $\lambda_{N\pi- N\pi}$ is replaced 
by a $N\to 2\pi N$ vertex amplitude for double-pion production.

The starting point for the construction of $\lambda_{N\pi- N\pi}$ is 
the phenomenological low-energy ($S$-wave) pion-nucleon 
interaction Hamiltonian ~\cite{KR}
\begin{eqnarray}
{\cal H}_{\rm{int}} = 
{\frac{\lambda_V}{m_\pi ^2}}
\bar{\psi}\gamma ^{\mu} \vegr{\tau} \psi \cdot
\left[\vegr{\phi} \times \partial_\mu \vegr{\phi}\right]
+ {\frac{\lambda_S}{m_\pi}}
\bar{\psi} \psi
\left[\vegr{\phi} \cdot \vegr{\phi}\right] .
\label{Lag-piN}
\end{eqnarray}

\sloppy
The calculations herein illustrated have been performed 
with the set of parameters used in the previous study~\cite{CantonSchad}.
Namely, for the isovector coupling constant in Eq.~(\ref{Lag-piN}), we have 
$\lambda_V/4{\pi} = 0.045$, whereas the $\pi NN$ and $\pi N\Delta$ coupling constants
are equal to ${f_{\pi NN}^2 / 4 \pi} = 0.0735$ 
and ${f_{\pi N\Delta}^2 / 4 \pi} = 0.32$, respectively. 
Our present treatment is based on Ref.~\cite{MWB_HO}.
The isospin-odd contribution is given in terms of 
a $\rho$-exchange model, while 
the isospin-even term is described as 
the combined effect of a phenomenological 
short-range (SR) process (Fig.~\ref{diag2}b) and an effective scalar-meson 
($\sigma$) exchange of Fig.~\ref{diag2}a. The two latter mechanisms act in opposite 
directions and almost cancel each other in case of the free pion-nucleon scattering. 
However, an off-shell enhancement of the probability amplitude takes place 
in the scalar-isoscalar channel for pion production off a bound system. 
A question still remains open, 
if this form mimics other physical effects, such as
the 2$N$ SR-exchange contributions proposed in Ref.~\cite{LeeRiska} (see also 
Ref.~\cite{Hanhart}, where the current status of theory and phenomenology 
of pion production in $NN$ collisions close to threshold is reviewed).

Further details (in particular, 
the treatment of the fully antisymmetrized matrix elements 
with respect to the Faddeev three-nucleon wave function) are not discussed 
here and can be found in the papers~\cite{CantonSchad,Canton_ea_1,Canton_ea_2}.
The three-body dynamics in the initial state has been described following
the treatment~\cite{Januschke} of the low-energy proton-deuteron
scattering, with the set of the AGS equations being reduced
through application of the spline-interpolation technique 
to a linear equation system.

\begin{figure}[hbtp]
  \begin{center}
    \resizebox{10cm}{!}{\includegraphics{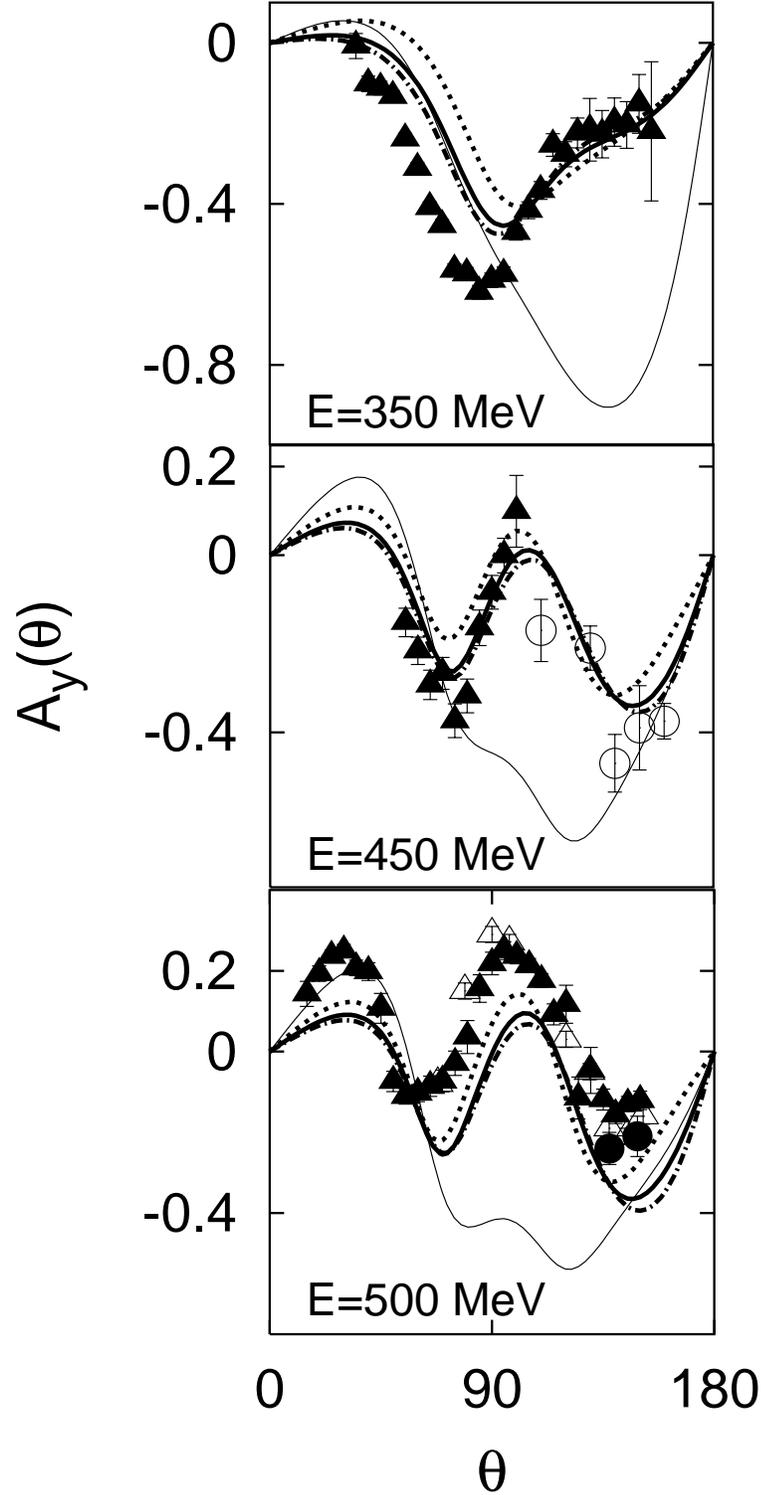}}
\caption{Proton analyzing power for three different
proton lab energies. Solid, dash-dotted and dotted lines denote our complete calculations
for Bonn-B, Bonn-A, and  Paris $NN$ potentials, respectively,  while 
thin solid line corresponds to the plane-wave calculation (without the $pd$ ISI). 
Full triangles, blank circles, blank 
triangles and full circles mark data of  
Refs.~\protect\cite{Cameron,Abegg,Cameron81} and~\protect\cite{Lolos}, correspondingly.}
    \label{fig-Ay-1}
  \end{center}
\end{figure}

\fussy
In Fig.~\ref{fig-Ay-1}, 
the results for the proton analyzing power 
are compared with what has been observed in experiments. For the three 
different phenomenological nucleon-nucleon potentials, one observes 
an additional peak at $\Theta_{\rm c.m.} \simeq 100^{\circ}$
when going from $E_{\rm lab}=350$ MeV towards higher energies.
This behaviour is fully explained by the model calculation, 
once the ISI involving the three-nucleon dynamics 
is taken into account. Otherwise, if one neglects these 
three-body effects in the initial channel, the structure of 
the observable is poorly reproduced.

\begin{figure}[hbtp]
  \begin{center}
    \resizebox{14cm}{!}{\includegraphics{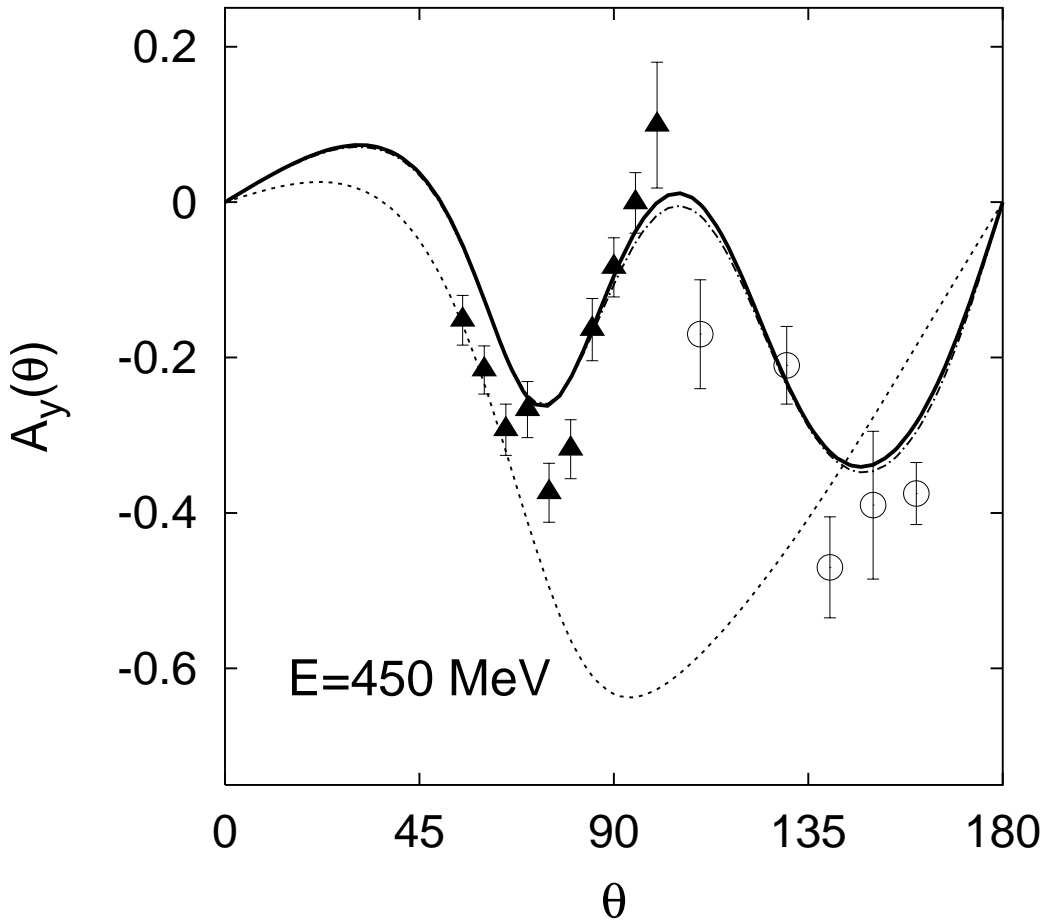}}
\caption{Results for different values of the maximum $\Delta N$ orbital momentum.
Dotted, solid and dashed-dotted curve refer to calculations with   
$l_\Delta=0$, $2$, and  $4$, respectively, 
performed with the Bonn-B potential and inclusion of the $3N$ ISI.}
    \label{fig-Ay-2}
  \end{center}
\end{figure}

In generation of this structure, another quantity that plays 
an important role is 
the angular momentum of the intermediate $\Delta N$ state. While $l_\Delta=0$
is sufficient in description of the general behaviour of 
the excitation function (see Ref.~\cite{CantonSchad}),
one must consider higher $\Delta N$ partial
waves, at least up to $l_{\Delta}=2$, in case of the analyzing power $A_y$. 
As follows from Fig.~\ref{fig-Ay-2}, 
there is no need, however, to go beyond this value, 
since the numerical convergence is already reached with the $\Delta N$ $D$-waves.

\begin{figure}[hbtp]
  \begin{center}
    \resizebox{16.4cm}{!}{\includegraphics{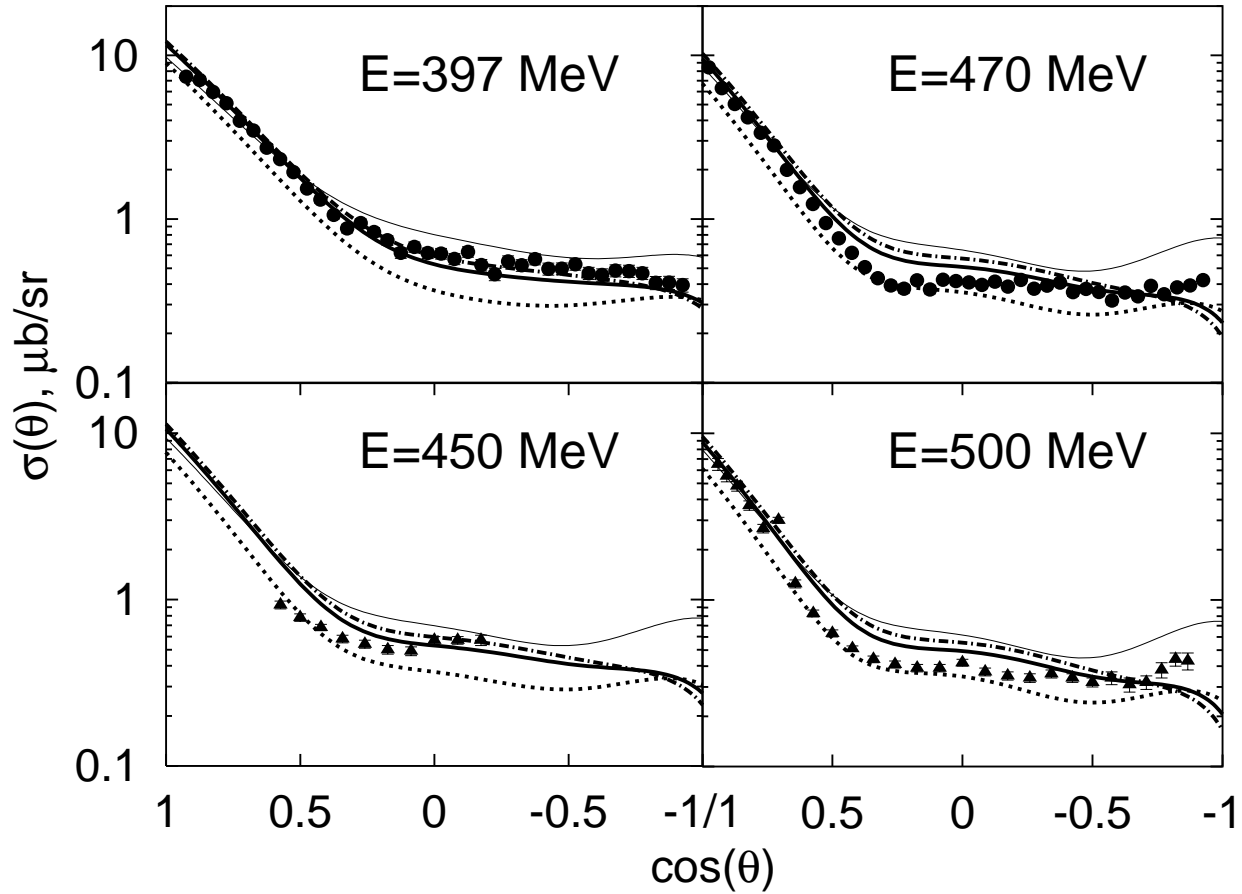}}
\caption{Differential cross-section 
of the ${p} d \rightarrow \pi^0 {\ ^{3}}\rm{He}$ reaction.
Data of Refs.~\protect\cite{GEM} and~\protect\cite{Cameron} 
are shown by full circles and triangles, respectively.
Notations for the curves are the same as in Fig.~\protect\ref{fig-Ay-1}.}
    \label{fig-diff-1}
  \end{center}
\end{figure}

Apart from the analyzing power $A_y$, also the angular 
differential cross-section 
has always been a challenge for the theory. 
Recent COSY data~\cite{GEM} provided new 
information about the shape of the cross-section, 
being at the same time consistent with earlier
TRIUMF measurements~\cite{Cameron}. 
All these data compare fairly well 
with our calculations, as shown in Fig.~\ref{fig-diff-1}.
Indeed, almost all of the data points fall in the strip determined
by the calculations performed with different $NN$ potentials.
It is worth noting that for higher energies and backward angles, 
the data seem to favor more the Paris potential, while
in the lower part of the energy range considered, 
the Bonn potentials (Bonn-B, in particular) give rise to 
the results which are closer to the data.

\sloppy
To summarize,
we have described the pionic capture processes in nucleon-deuteron
collisions employing elementary production mechanisms obtained from 
the non-relativistic reduction of the phenomenological $\pi N$ 
and $\pi\Delta$ Hamiltonians. The corresponding matrix elements
between the three-nucleon states have been evaluated within a 
large-basis space, consisting of 464 three-nucleon partial waves.
Total angular momenta of the system up to $7/2$ for both the parities
have been considered. The prescriptions required by the Pauli principle 
have been taken into account through the application of the
permutation formalism to the three-nucleon system. Finally,
these matrix elements have been folded with the nuclear WF to 
obtain the probability amplitudes, with the complete details
of the procedure being outlined in Refs.~\cite{CantonSchad,Canton_ea_1,Canton_ea_2}.
In the calculation of the plane-wave part of the amplitude, we mainly 
follow these papers. The only difference is modification of the standard 
Jacobi momentum set for the system through the replacement of 
the pion mass by its total energy, what is more adequate to 
the kinematics of the present calculations. 
An extended paper explaining this and some other ingredients of 
the model is in progress. 
However, much more important turns out to be our regard of the $3N$ 
dynamics in the initial proton-deuteron state, which was not extensively 
investigated in the previous treatment of the reaction. 

\fussy
The results obtained suggest that one achieves a fairly good reproduction
of the phenomenology, when including, 
in addition to the $\Delta$-rescattering mechanism, the 
$\rho$-exchange process in the $S$-wave, as well as
another $S$-wave mechanism that generates a significant amount of interaction
in the scalar-isoscalar $\pi N$ channel. 
On the contrary, the 
one-body term (the off-energy-shell ${\pi { NN}}$ vertex) adds contributions 
that are significantly smaller. For the scalar-isoscalar component
we have considered an off-shell model~\cite{MWB_HO}, which has been used
before. 
We do not claim here that this particular way to generate the flux in
the scalar-isoscalar channel is more realistic than other 
alternatives ({e.g.}, the above-mentioned SR-exchange 
contributions proposed in Ref.~\cite{LeeRiska}). 
A study on how really the flux is generated in the scalar-isoscalar channel 
is beyond the scope of the present work, but can  
represent at this point a subject for future investigations. 
We hope that studies of spin observables close to threshold
of this specific reaction can shed light on that problem.

This work shows that, with a combination of two-body
mechanisms, one can explain the structure of the 
proton analyzing power $A_y$,
which has been accurately measured about 20 years ago
and represented since then a challenge for theorists, 
being a difficult observable to interpret.
We conclude that it is possible to reproduce qualitatively the 
nontrivial energy dependence of $A_y$ around the $\Delta$ resonance, 
provided that the three-body 
nuclear dynamics in the initial state is taken into account. 
It is worth mentioning also that we have made use of the techniques 
that have been developed, in fact, mainly for 
low-energy few-nucleon applications. It is remarkable that they can be 
successfully used in this energy regime, where the use of the three-body
non-relativistic quantum-mechanical equations and the concept of the 
standard $NN$ potentials is, in some sense, questionable. 

\bigskip

L.G.L. is indebted to the University of Padova and INFN for 
their kind hospitality 
and support in October-November, 2003 and April-August, 2004.

\newcommand{\etal}{{\em et al.}}


\end{document}